\documentclass[pra,twocolumn,showpacs,draft]{revtex4}
\usepackage{amssymb,amsmath,amssymb}
\usepackage{times}
\usepackage{bbm,bm}
\newcommand{\p}{{\bm r}}
\newcommand{\s}{{\bm \sigma}}
\newcommand{\tr}{\mathrm{tr}}
\newcommand{\ra}{\rangle}
\newcommand{\la}{\langle}
\newcommand{\sv}{{\bm s}}
\newcommand{\nv}{{\bm n}}
\newcommand{\iv}{{\bm i}}
\newcommand{\mv}{{\bm m}}
\newcommand{\sq}{\varrho}
\newcommand{\lm}{\Lambda_{\max}^2}

\begin{document}

\title{Analytic Expressions for Geometric Measure of Three Qubit States}

\author{Levon Tamaryan}

\affiliation{Physics Department, Yerevan State University,
Yerevan, 375025, Armenia}

\email{levtam@mail.yerphi.am}

\author{DaeKil Park}

\affiliation{Department of Physics, Kyungnam University, Masan,
631-701, Korea}

\email{dkpark@hep.kyungnam.ac.kr}

\author{Sayatnova Tamaryan}

\affiliation{Theory Department, Yerevan Physics Institute,
Yerevan, 375036, Armenia}

\email{sayat@mail.yerphi.am}


\begin{abstract}
A new method is developed to derive an algebraic equations for the
geometric measure of entanglement of three qubit pure states. The equations
are derived explicitly and solved in cases of most interest. These
equations allow oneself to derive the analytic expressions of the
geometric entanglement measure in the wide range of the three
qubit systems, including the general class of W-states and states
which are symmetric under permutation of two qubits. The nearest
separable states are not necessarily unique and highly entangled
states are surrounded by the one-parametric set of equally distant
separable states. A possibility for the physical applications of
the various three qubit states to quantum teleportation and
superdense coding is suggested from the aspect of the
entanglement.
\end{abstract}

\maketitle

\section{Introduction.}

Entangled states have different remarkable applications and among
them are quantum cryptography \cite{ek,exp-cr}, superdense coding
\cite{dence,exp-den}, teleportation \cite{bentel,exp-tel} and the
potential speedup of quantum algorithms
\cite{speed,vid,Shim-shor}. The entanglement of bipartite systems
is well-understood \cite{Ben,benn,woot,niels}, while the
entanglement of multipartite systems offers a real challenge to
physicists. In contrast to bipartite setting, there is no unique
treatment of the maximally entangled states for multipartite
systems. In this reason it is highly difficult to formulate a
theory of multipartite entanglement. Another point which makes
difficult to understand the entanglement for the multi-qubit
systems is mainly due to the fact that the analytic expressions
for the various entanglement measures is extremely hard to derive.

We consider pure three qubit systems \cite{acin,lind,coff,red},
although the entanglement of mixed states attracts a considerable
attention. For example, in recent experiment \cite{swiz} the
tangle for general mixed states was evaluated, which has never
been done before. Three-qubit system is important in the sense
that it is the simplest system which gives a non-trivial effect in
the entanglement. Thus, we should understand the general
properties of the entanglement in this system as much as possible
to go further more complicated higher qubit system. The
three-qubit system can be entangled in two inequivalent ways GHZ
\cite{ghz} and W, and neither form can be transformed into the
other with any probability of success \cite{Chir}. This picture is
complete: any fully entangled state is SLOCC equivalent to either
GHZ or W.

Only very few analytical results for tripartite entanglement have
been obtained so far \cite{iso} and we need more light on the
subject. This is our main objective and we choose geometric
measure of entanglement $E_g$ \cite{vedr,Shim,barn,wei}. It is an
axiomatic measure \cite{vedr,pop,vp,hor}, is connected with other
measures \cite{con1,con2} and has an operational treatment.
Namely, for the case of pure states it is closely related to the
Groverian measure of entanglement \cite{bno} and the latter is
associated with the success probability of Grover's search
algorithm \cite{grov} when a given state is used as the initial
state.

Geometric measure depends on entanglement eigenvalue $\lm$ and is
given by formula $E_g(\psi)=1-\lm$. For pure states the
entanglement eigenvalue is equal to the maximal overlap of a given
state with any complete product state. The maximization over
product states gives {\it nonlinear eigenproblem} \cite{wei}
which, except rare cases, does not allow the complete analytical
solutions.

Recently the idea was suggested that nonlinear eigenproblem can be
reduced to the linear eigenproblem for the case of three qubit
pure states \cite{we}. The idea is based on theorem stating that
any reduced $(n-1)$-qubit state uniquely determines the geometric
measure of the original $n$-qubit pure state. This means that two
qubit mixed states can be used to calculate the geometric measure
of three qubit pure states and this will be fully addressed in
this work.

The method gives two algebraic equations of degree six defining
the geometric measure of entanglement. Thus the difficult problem
of geometric measure calculation is reduced to the algebraic
equation root finding. Equations contain valuable information, are
good bases for the numerical calculations and may test numerical
calculations based on other numerical techniques \cite{Shim-shor}.

Furthermore, the method allows to find the nearest separable
states for three qubit states of most interest and get analytic
expressions for their geometric measures. It turn out that highly
entangled states have their own feature. Each highly entangled
state has a vicinity with no product state and all nearest product
states are on the boundary of the vicinity and form an
one-parametric set.

In Section II we derive algebraic equations defining the geometric
entanglement measure of pure three qubit states and present the
general solution. In Section III we examine W-type states and
deduce analytic expression for their geometric measures. States
symmetric under permutation of two qubits are considered in
Section IV, where the overlap of the state functions with the
product states are maximized directly. In last Section V we make
concluding remarks.

\section{Algebraic equations.}

We consider three qubits A,B,C with state function $|\psi\ra$. The
entanglement eigenvalue is given by

\begin{equation}\label{gen.lm}
\Lambda_{max}=\max_{q^1q^2q^3}|\la q^1q^2q^3|\psi\ra|
\end{equation}

\noindent and the maximization runs over all normalized complete
product states $|q^1\ra\otimes|q^2\ra\otimes|q^3\ra$. Superscripts
label single qubit states and spin indices are omitted for
simplicity. Since in the following we will use density matrices
rather than state functions, our first aim is to rewrite
Eq.(\ref{gen.lm}) in terms of density matrices. Let us denote by
$\rho^{ABC}=|\psi\ra\la\psi|$ the density matrix of the
three-qubit state and by $\sq^k=|q^k\ra\la q^k|$ the density
matrices of the single qubit states. The equation for the square
of the entanglement eigenvalue takes the form

\begin{equation}\label{gen.pmax}
\lm(\psi)=\max_{\sq^1\sq^2\sq^3}
\tr\left(\rho^{ABC}\sq^1\otimes\sq^2\otimes\sq^3\right).
\end{equation}

An important equality

\begin{equation}\label{gen.unit}
\max_{\sq^3} \tr(\rho^{ABC}\sq^1\otimes\sq^2\otimes\sq^3)=
\tr(\rho^{ABC}\sq^1\otimes\sq^2\otimes\openone^3)
\end{equation}

\noindent was derived in \cite{we} where $\openone$ is a unit
matrix. It has a clear meaning. The matrix
$\tr(\rho^{ABC}\sq^1\otimes\sq^2)$ is $2\otimes2$ hermitian matrix
and has two eigenvalues. One of eigenvalues is always zero and
another is always positive and therefore the maximization of the
matrix simply takes the nonzero eigenvalue. Note that its minimization
gives zero as the minimization takes the zero eigenvalue.

We use Eq.(\ref{gen.unit}) to reexpress the entanglement
eigenvalue by reduced density matrix $\rho^{AB}$ of qubits A and B
in a form

\begin{equation}\label{gen.pred}
\lm(\psi)=\max_{\sq^1\sq^2}
\tr\left(\rho^{AB}\sq^1\otimes\sq^2\right).
\end{equation}

We denote by $\sv_1$ and $\sv_2$ the unit Bloch vectors of
the density matrices $\sq^1$ and $\sq^2$ respectively and adopt
the usual summation convention on repeated indices $i$ and $j$.
Then

\begin{equation}\label{gen.s1s2}
\lm=\frac{1}{4}\max_{s_1^2=s_2^2=1}\left(1+\sv_1\cdot
\p_1+\sv_2\cdot \p_2+g_{ij}\,s_{1i}s_{2j}\right),
\end{equation}

where

\begin{equation}\label{gen.vec}
\p_1=\tr(\rho^A\s),\,\,\p_2=\tr(\rho^B\s),\,\,
g_{ij}=\tr(\rho^{AB}\sigma_i\otimes\sigma_j)
\end{equation}

\noindent and $\sigma_i$'s are Pauli matrices. The matrix $g_{ij}$
is not necessarily to be symmetric but must has only real entries. The
maximization gives a pair of equations

\begin{equation}\label{gen.eq}
\p_1+g\sv_2=\lambda_1\sv_1,\quad\p_2+g^T\sv_1=\lambda_2\sv_2,
\end{equation}

\noindent where Lagrange multipliers $\lambda_1$ and $\lambda_2$
are enforcing unit nature of the Bloch vectors. The solution of Eq.(\ref{gen.eq})
is

\begin{subequations}\label{gen.sol}
\begin{equation}\label{gen.sol1}
\sv_1=\left(\lambda_1\lambda_2\openone-g\,g^T\right)^{-1}
\left(\lambda_2\p_1+g\,\p_2\right),
\end{equation}
\begin{equation}\label{gen.sol2}
\sv_2=\left(\lambda_1\lambda_2\openone-g^Tg\right)^{-1}
\left(\lambda_1\p_2+g^T\p_1\right).
\end{equation}
\end{subequations}

\noindent Now, the only unknowns are Lagrange multipliers, which should be
determined by equations

\begin{equation}\label{gen.alg}
|\sv_1|^2=1,\quad |\sv_2|^2=1.
\end{equation}

In general, Eq.(\ref{gen.alg}) give two algebraic equations of
degree six. However, the solution (\ref{gen.sol}) is valid if
Eq.(\ref{gen.eq}) supports a unique solution and this is by no
means always the case. If the solution of Eq.(\ref{gen.eq})
contains a free parameter, then
$\det(\lambda_1\lambda_2\openone-gg^T)=0$ and, as a result,
Eq.(\ref{gen.sol}) cannot not applicable. The example presented in
Section III will demonstrate this situation.

In order to test Eq.(\ref{gen.sol}) let us consider an arbitrary
superposition of W

\begin{equation}\label{ww.w}
|W\ra=\frac{1}{\sqrt{3}}\left(|100\ra+|010\ra+|001\ra\right)
\end{equation}

\noindent and flipped W

\begin{equation}\label{ww.flip}
|\widetilde{W}\ra=\frac{1}{\sqrt{3}}\left(|011\ra+|101\ra+|110\ra\right)
\end{equation}

\noindent states, i.e. the state

\begin{equation}
|\psi\ra=\cos\theta|W\ra+\sin\theta|\widetilde{W}\ra.
\end{equation}

Straightforward calculation yields

\begin{subequations}\label{ww.matr}
\begin{equation}\label{ww.matrv}
\p_1=\p_2=\frac{1}{3}\left(2\sin2\theta\iv+\cos2\theta\nv\right),
\end{equation}
\begin{equation}\label{ww.matrm}
g=\frac{1}{3}
\begin{pmatrix}
2 & 0 & 0\\
0 & 2 & 0\\
0 & 0 &\!\! -1
\end{pmatrix}
,
\end{equation}
\end{subequations}

\noindent where unit vectors $\iv$ and $\nv$ are aligned with the
axes $x$ and $z$, respectively. Both vectors $\iv$ and $\nv$ are
eigenvectors of matrices $g$ and $g^T$. Therefore $\sv_1$ and
$\sv_2$ are linear combinations of $\iv$ and $\nv$. Also from
$\p_1=\p_2$ and $g=g^T$ it follows that $\sv_1=\sv_2$ and
$\lambda_1=\lambda_2$. Then Eq.(\ref{gen.sol}) for general
solution give

\begin{equation}\label{ww.sv}
\sv_1=\sv_2=\sin2\varphi\,\iv+ \cos2\varphi\,\nv
\end{equation}

\noindent where

\begin{equation}\label{ww.var}
\sin2\varphi=\frac{2\sin2\theta}{3\lambda-2},\quad
\cos2\varphi=\frac{\cos2\theta}{3\lambda+1}.
\end{equation}

The elimination of the Lagrange multiplier $\lambda$ from
Eq.(\ref{ww.var}) gives

\begin{equation}\label{ww.el}
3\sin2\varphi\cos2\varphi=\cos2\theta\sin2\varphi-
2\sin2\theta\cos2\varphi.
\end{equation}

Let us denote by $t=\tan\varphi$. After the separation of the
irrelevant root $t=-\tan\theta$, Eq.(\ref{ww.el}) takes the form

\begin{equation}\label{ww.eq}
\sin\theta\,t^3+2\cos\theta\,t^2-2\sin\theta\,t-\cos\theta=0.
\end{equation}

\noindent This equation exactly coincides with that derived in
\cite{wei}. Since a detailed analysis was given in Ref.\cite{wei},
we do not want to repeat the same calculation here.
Instead we would like to consider the three-qubit states that allow
the analytic expressions for the geometric entanglement measure by making use
of Eq.(\ref{gen.eq}).

\section{W-type states.}

Consider W-type state

\begin{equation}\label{w.psi}
|\psi\ra=a|100\ra+b|010\ra+c|001\ra,\quad a^2+b^2+c^2=1.
\end{equation}

\noindent Without loss of generality we  consider only the case of
positive parameters $a,b,c$. Direct calculation yields

\begin{equation}\label{w.matr}
 \p_1=r_1\,\nv,\quad\p_2=r_2\,\nv,\quad g=
\begin{pmatrix}
\omega & 0 & 0\\
0 & \omega & 0\\
0 & 0 & -r_3
\end{pmatrix}
,
\end{equation}
where
\begin{equation}\label{w.eig}
r_1=b^2+c^2-a^2,\, r_2=a^2+c^2-b^2,\,r_3=a^2+b^2-c^2
\end{equation}

\noindent and $\omega=2ab$. The unit vector $\nv$ is  aligned with
the axis $z$. Any vector perpendicular to $\nv$ is an eigenvector
of $g$ with eigenvalue $\omega$. Then from Eq.(\ref{gen.eq}) it
follows that the components of vectors $\sv_1$ and $\sv_2$
perpendicular to $\nv$ are collinear. We denote by $\mv$ the unit
vector along that direction and parameterize vectors $\sv_1$ and
$\sv_2$ as follows

\begin{equation}\label{w.par}
\sv_1=\cos\alpha\,\nv+\sin\alpha\,\mv,\quad
\sv_2=\cos\beta\,\nv+\sin\beta\,\mv.
\end{equation}

Then Eq.(\ref{gen.eq}) reduces to the following four equations

\begin{subequations}\label{w.angle}
\begin{equation}\label{w.cos}
r_1-r_3\cos\beta=\lambda_1\cos\alpha,\quad
r_2-r_3\cos\alpha=\lambda_2\cos\beta,
\end{equation}
\begin{equation}\label{w.sin}
\omega\sin\beta=\lambda_1\sin\alpha,\quad
\omega\sin\alpha=\lambda_2\sin\beta,
\end{equation}
\end{subequations}

\noindent which are used to solve the four unknown constants
$\lambda_1,\lambda_2,\alpha$ and $\beta$.
Eq.(\ref{w.sin}) impose either

\begin{equation}\label{w.l1l2}
\lambda_1\lambda_2-\omega^2=0
\end{equation}

\noindent or

\begin{equation}\label{w.sinsin}
\sin\alpha\sin\beta=0.
\end{equation}

First consider the case $r_1>0,r_2>0,r_3>0$ and coefficients
$a,b,c$ form an acute triangle. Eq.(\ref{w.sinsin}) does not give
a true maximum and this can be understood as follows. If both
vectors $\sv_1$ and $\sv_2$ are aligned with the axis $z$, then
the last term in Eq.(\ref{gen.s1s2}) is negative. If vectors
$\sv_1$ and $\sv_2$ are antiparallel, then one of scalar products
in Eq.(\ref{gen.s1s2}) is negative. In this reason $\lm$ cannot be
maximal. Then Eq.(\ref{w.l1l2}) gives true maximum and we have to
choose positive values for $\lambda_1$ and $\lambda_2$ to get
maximum.

First we use Eq.(\ref{w.cos}) to connect the angles  $\alpha$ and
$\beta$ with the Lagrange multipliers $\lambda_1$ and $\lambda_2$

\begin{equation}\label{w.cosval}
\cos\alpha=\frac{\lambda_2r_1-r_2r_3}{\omega^2-r_3^2},\quad
\cos\beta=\frac{\lambda_1r_2-r_1r_3}{\omega^2-r_3^2}.
\end{equation}

Then Eq.(\ref{w.sin}) and (\ref{w.l1l2}) give the following
expressions for Lagrange multipliers $\lambda_1$ and $\lambda_2$

\begin{subequations}\label{w.lagroot}
\begin{equation}\label{w.lagroot1}
\lambda_1=\omega\left(\frac{\omega^2+r_1^2-r_3^2}
{\omega^2+r_2^2-r_3^2}\right)^{1/2},
\end{equation}
\begin{equation}\label{w.lagroot2}
\lambda_2=\omega\left(\frac{\omega^2+r_2^2-r_3^2}
{\omega^2+r_1^2-r_3^2}\right)^{1/2}.
\end{equation}
\end{subequations}

Eq.(\ref{gen.eq}) allows to write a shorter expression for the
entanglement eigenvalue

\begin{equation}\label{w.lsimple}
\lm=\frac{1}{4}\left(1+\lambda_2+r_1\cos\alpha\right).
\end{equation}

Now we insert the values of $\lambda_2$ and $\cos\alpha$ into
Eq.(\ref{w.lsimple}) and obtain

\begin{equation}\label{w.frac}
4\lm=1+\frac{\omega\sqrt{(\omega^2+r_1^2-r_3^2)
(\omega^2+r_2^2-r_3^2)}-r_1r_2r_3}{\omega^2-r_3^2}.
\end{equation}

The denominator in above expression is  multiple of the area $S$
of the triangle $a,b,c$

\begin{equation}\label{w.area}
\omega^2-r_3^2=16S^2.
\end{equation}

A little algebra yields for the numerator

\begin{eqnarray}\label{w.numer}
 & & \omega\sqrt{(\omega^2+r_1^2-r_3^2)+
(\omega^2+r_2^2-r_3^2)}-r_1r_2r_3    \\ \nonumber
& &\hspace{.3cm} = 16\,a^2b^2c^2-\omega^2+r_3^2.
\end{eqnarray}

Combining together the numerator and denominator, we obtain the
final expression for the entanglement eigenvalue

\begin{equation}\label{w.rad}
\lm=4R^2,
\end{equation}

\noindent where $R$ is the circumradius of the triangle $a,b,c$.
Entanglement value is minimal  when triangle is regular, i.e. for
W-state and $\lm(W)=4/9$ \cite{Shim-grov,wei}.

Now consider the case $r_3<0$. Since $r_3+r_1=2b^2\geq0$,
we have $r_1>0$ and similarly $r_2>0$. Eq.(\ref{w.sinsin})
gives true maximum in this case and both vectors are aligned with
the axis $z$

\begin{equation}\label{w.parz}
\sv_1=\sv_2=\nv
\end{equation}

\noindent resulting in $\lm=c^2$. In view of symmetry

\begin{equation}\label{w.max}
\lm=\max(a^2,b^2,c^2),\quad \max(a^2,b^2,c^2)>\frac{1}{2}.
\end{equation}

Since the matrix $g$ and vectors $\p_1$ and $\p_2$ are invariant
under rotations around axis $z$ the same properties must have
Bloch vectors $\sv_1$ and $\sv_2$. There are two possibilities:

\medskip

i)Bloch vectors are unique and aligned with the axis $z$. The
solution given by Eq.(\ref{w.parz}) corresponds to this situation
and the resulting entanglement eigenvalue Eq.(\ref{w.max})
satisfies the inequality

\begin{equation}\label{w.slight}
\frac{1}{2}<\lm\leq1.
\end{equation}

ii)Bloch vectors have nonzero components in $xy$ plane and the
solution is not unique.  Eq.(\ref{w.par}) corresponds to this
situation and contains a free parameter. The free parameter is the
angle defining the direction of the vector $\mv$ in the $xy$
plane. Then Eq.(\ref{w.rad}) gives the entanglement eigenvalue in
highly entangled region

\begin{equation}\label{w.high}
\frac{4}{9}\leq\lm<\frac{1}{2}.
\end{equation}

Eq.(\ref{w.rad}) and (\ref{w.max}) have joint curves when
parameters $a,b,c$ form a right triangle and give $\lm=1/2$. The
GHZ states have same entanglement value and it seems to imply
something interesting. GHZ state can be used for teleportation and
superdense coding, but W-state cannot be. However, the W-type
state with right triangle coefficients can be used for
teleportation and superdense coding \cite{agr}. In other words,
both type of states can be applied provided they have the required
entanglement eigenvalue $\lm=1/2$.

\section{Symmetric States.}

Now let us consider the state which is symmetric under permutation of
qubits A and B and contains three real independent parameters

\begin{equation}\label{ghz.psi}
|\psi\ra=a|000\ra+b|111\ra+c|001\ra+d|110\ra,
\end{equation}

\noindent where $a^2+b^2+c^2+d^2=1$. According to Generalized
Schmidt Decomposition \cite{acin} the states with different sets
of parameters are local-unitary(LU) inequivalent. The relevant quantities are

\begin{equation}\label{ghz.matr}
 \p_1=\p_2=r\,\nv,\quad g=
\begin{pmatrix}
\omega & 0 & 0\\
0 & -\omega & 0\\
0 & 0 & 1
\end{pmatrix}
,
\end{equation}

\noindent where

\begin{equation}\label{ghz.eig}
r=a^2+c^2-b^2-d^2,\quad \omega=2ad+2bc
\end{equation}

\noindent and the unit vector $\nv$ again is  aligned with the
axis $z$.

All three terms in the l.h.s. of Eq.(\ref{gen.s1s2}) are bounded
above:
\begin{itemize}

\item $\sv_1\cdot\p_1\leq|r|$,

\item $\sv_2\cdot\p_2\leq|r|$,

\item and owing to inequality $|\omega|\leq1,
\,g_{ij}\,s_{1i}s_{2j}\leq1$.

\end{itemize}

Quite surprisingly all upper limits are reached simultaneously at

\begin{equation}\label{ghz.sign}
\sv_1=\sv_2={\rm Sign}(r)\nv,
\end{equation}

\noindent which results in

\begin{equation}\label{ghz.mod}
\lm=\frac{1}{2}\left(1+|r|\right).
\end{equation}

This expression has a clear meaning. To understand it we
parameterize the state as

\begin{equation}\label{ghz.repsi}
|\psi\ra=k_1|00q_1\ra+k_2|11q_2\ra,
\end{equation}

\noindent where $q_1$ and $q_2$ are arbitrary single normalized
qubit states and positive parameters $k_1$ and $k_2$ satisfy
$k_1^2+k_2^2=1$. Then

\begin{equation}\label{ghz.l}
\lm=\max(k_1^2,k_2^2),
\end{equation}

\noindent i.e. the maximization takes a larger coefficient in
Eq.(\ref{ghz.repsi}). In bipartite case the maximization takes the
largest coefficient in Schmidt decomposition \cite{vidjon,bno} and
in this sense Eq.(\ref{ghz.repsi}) effectively takes the place of
Schmidt decomposition. When $|q_1\ra=|0\ra$  and $|q_2\ra=|1\ra$, Eq.(\ref{ghz.l})
gives the known answer for generalized GHZ state
\cite{Shim-grov,wei}.

The entanglement eigenvalue is minimal $\lm=1/2$ on condition that
$k_1=k_2$. These states can be described as follows

\begin{equation}\label{ghz.half}
|\psi\ra=|00q_1\ra+|11q_2\ra
\end{equation}

\noindent where $q_1$ and $q_2$ are arbitrary single qubit
normalized states. The entanglement eigenvalue is constant
$\lm=1/2$ and does not depend on single qubit state parameters.
Hence one may expect that all these states can be applied for
teleportation and superdense coding. It would be interesting to
check whether this assumption is correct or not.

It turns out that GHZ state is not a unique state and is one of
two-parametric LU inequivalent states that have $\lm=1/2$.
On the other hand W-state is unique up to LU transformations and the low
bound $\lm=4/9$ is reached if and only if $a=b=c$. However, one
cannot make such conclusions in general. Five real parameters are
necessary to parameterize the set of inequivalent three qubit pure
states \cite{acin}. And there is no explicit argument that W-state
is not just one of LU inequivalent states that have $\lm=4/9$.

\section{Summary.}

We have derived algebraic equations defining geometric measure of
three qubit pure states. These equations have a degree higher than
four and explicit solutions for general cases cannot be derived
analytically. However, the explicit expressions are not important.
Remember that explicit expressions for the algebraic equations of
degree three and four have a limited practical significance but
the equations itself are more important. This is especially true
for equations of higher degree; main results can be derived from
the equations rather than from the expressions of their roots.

Eq.(\ref{gen.eq}) give the nearest separable state directly and
this separable states have useful applications. In order to
construct an entanglement witness, for example, the crucial point lies
in finding the nearest separable state \cite{bert}. This will be
especially interesting for highly entangled states that have a
whole set of nearest separable states and allow to construct a set
of entanglement witnesses.

The expression in r.h.s. of Eq.(\ref{gen.s1s2}) can be maximized
directly for various three qubit states. Although it is very hard
to solve the higher-degree equation, it turns out that the wide
range of the three-qubit states have a symmetry and this symmetry
reduces the equations of degree six to the quadratic equations. In
this reason Eq.(\ref{gen.s1s2}) can be used to derive the analytic
expressions of the various entanglement measures for the
three-qubit states. Also Eq.(\ref{gen.s1s2}) can be a starting
point to explore the numerical computation of the entanglement
measures for the higher-qubit systems. We would like to discuss
this issue elsewhere.

\bigskip

\begin{acknowledgments}
LT is grateful to Roland Avagyan for help. ST thanks Jin-Woo Son,
Eylee Jung, Mi-Ra Hwang, Hungsoo Kim and Min-Soo Kim for
illuminating conversations. DKP was supported by the Kyungnam
University Research Fund, 2007.
\end{acknowledgments}


\begin{thebibliography}{99}
\bibitem{ek}A. K. Ekert, Phys. Rev. Lett. {\bf67}, 661 (1991).
\bibitem{exp-cr}C. H. Bennett, F. Bessette, G. Brassard, L. Salvail and
J. Smolin, J. Cryptology {\bf5}, 3 (1992).
\bibitem{dence}C. H. Bennett and S. J. Wiesner, Phys. Rev. Lett. {\bf
69}, 2881 (1992).
\bibitem{exp-den}K. Mattle, H. Weinfurter, P. G. Kwiat and
A. Zeilinger, Phys. Rev. Lett. {\bf76}, 4656 (1996).
\bibitem{bentel}C. H. Bennett, G. Brassard, C. Crepeau, R. Jozsa,
A. Peres, W. K. Wootters, Phys. Rev. Lett. {\bf 70}, 1895 (1993).
\bibitem{exp-tel}D. Boschi, S. Branca, F. De Martini, L. Hardy and
S. Popescu, Phys. Rev. Lett. {\bf80}, 1121 (1998).
\bibitem{speed}R. Jozsa and N. Linden, Proc. R. Soc. London, A{\bf459},
2011 (2003).
\bibitem{vid}G. Vidal, Phys. Rev. Lett. {\bf 91}, 147902
(2003).
\bibitem{Shim-shor}Y. Shimoni, D. Shapira, and O. Biham, Phys. Rev. A {\bf
72}, 062308 (2005).
\bibitem{Ben}C. H. Bennett, H. J. Bernstein, S. Popescu and
B. Schumacher, Phys.Rev. A {\bf 53}, 2046 (1996).
\bibitem{benn}C. H. Bennett, D. P. DiVincenzo, J. Smolin and
W. K. Wootters, Phys. Rev. A {\bf 54}, 3824(1997).
\bibitem{woot}W. K. Wootters, Phys. Rev. Lett. {\bf 80}, 2245 (1998).
\bibitem{niels}M. A. Nielsen, Phys. Rev. Lett. {\bf 83}, 436 (1999).
\bibitem{acin}A. Acin, A. Andrianov, E. Jane, J. I. Latorre, R. Tarrach,
Phys. Rev. Lett. {\bf 85}, 7 (2000).
\bibitem{lind}N. Linden, S. Popescu and A. Sudbery, Phys. Rev. Lett.
{\bf 83}, 243 (1999).
\bibitem{coff}V. Coffman, J. Kundu and W. K. Wootters, Phys. Rev. A{\bf
61}, 052306 (2000).
\bibitem{red}N. Linden, S. Popescu and W. K. Wootters, Phys. Rev.
Lett. {\bf 89}, 207901 (2002).
\bibitem{swiz}B. R\"othlisberger, J. Lehmann, D. S. Saraga, Ph. Traber
and D. Loss,     arXiv:0705.1710v1 [quant-ph].
\bibitem{ghz}D. Greenberger, M. Horne and A. Zeilinger, Phys. Today,
August 1993, 24.
\bibitem{Chir}W. D\"{ur}, G. Vidal and J. I. Cirac, Phys. Rev. A {\bf
62}, 062314 (2000).
\bibitem{iso}B. M. Terhal and K. G. H. Vollbrecht, Phys. Rev. Lett. {\bf
85}, 2625 (2000).
\bibitem{vedr}V. Vedral, M. B. Plenio, M. A. Rippin, P. L. Knight,
Phys. Rev. Lett. {\bf 78}, 2275 (1997).
\bibitem{Shim}A. Shimony, Ann. NY. Acad. Sci {\bf 755}, 675 (1995).
\bibitem{barn}H. Barnum and N. Linden, J. Phys. A: Math. Gen. {\bf 34},
(2001) p.6787.
\bibitem{wei}T.-C. Wei and P. M. Goldbart, Phys. Rev. A {\bf 68},
042307 (2003).
\bibitem{pop}S. Popescu and D. Rohrlich, Phys. Rev. A {\bf 56},
3219(1997).
\bibitem{vp}V. Vedral and M. Plenio, Phys. Rev. A {\bf 57}, 1619
(1998).
\bibitem{hor}M. Horodecki, P. Horodecki and R. Horodecki,
Phys. Rev. Lett. {\bf 84}, 2014(2000).
\bibitem{con1}T.-C. Wei, M. Ericsson, P. M. Goldbart and W. J. Munro,
Quant. Inf. Comp. {\bf 4}, 252 (2004).
\bibitem{con2}D. Cavalcanti, Phys. Rev. A {\bf 73}, 044302 (2006).
\bibitem{bno}O. Biham, M. A. Nielsen and T. J. Osborne,
Phys. Rev. A {\bf65}, 062312 (2002).
\bibitem{grov}L. K. Grover, Phys. Rev. Lett. {\bf 79}, 325 (1997).
\bibitem{we}E. Jung, Mi-Ra Hwang, H. Kim, M.-S. Kim,
D. K. Park, J.-W. Son and S. Tamaryan, arXiv:0709.4292v1
[quant-ph]
\bibitem{Shim-grov}Y. Shimoni, D. Shapira, and O. Biham, Phys. Rev.
A {\bf 69}, 062303 (2004).
\bibitem{agr}P. Agrawal and A. Pati, Phys. Rev. A {\bf 74}, 062320
(2006).
\bibitem{vidjon}G. Vidal, D. Jonathan, and M. A. Nielsen, Phys.
Rev. A {\bf 62}, 012304 {2000}.
\bibitem{bert}R. A. Bertlmann and Ph. Krammer, arXiv: 0710.1184v1
[quant-ph].
\end{thebibliography}
\end{document}